\documentclass[floatfix,aps,pre,twocolumn,nofootinbib,superscriptaddress]{revtex4}

\usepackage{graphicx}
\usepackage{epsfig}

\usepackage{amsmath}
\usepackage{amsfonts}
\usepackage{amssymb}
\usepackage{bm}
\usepackage{mathtools}
\usepackage{comment}
\usepackage{cancel}

\usepackage{hyperref}
\usepackage{dsfont}
\usepackage{lipsum}

\usepackage{tikz}
\usepackage{circuitikz}
\usetikzlibrary{arrows, shapes}
\usepackage{pgfplots}
\pgfplotsset{compat=1.14}

\usepackage{pdfpages}

\newcommand{\mean}[1]{\left\langle #1 \right\rangle}
\newcommand{\meann}[1]{\langle #1 \rangle}

\begin{document}
\title{Bridging Freidlin-Wentzell large deviations theory and stochastic thermodynamics}
\author{Davide Santolin}
\affiliation{Department of Physics and Astronomy,
University of Padova,
Via Marzolo 8, I-35131 Padova,
Italy}
\author{Nahuel Freitas}
\affiliation{Departamento de Física, FCEyN, UBA, Pabellón 1, Ciudad Universitaria, 1428 Buenos Aires, Argentina}
\author{Massimiliano Esposito}
\affiliation{Complex Systems and Statistical Mechanics, Department of Physics and Materials Science,
University of Luxembourg, L-1511 Luxembourg, Luxembourg}
\author{Gianmaria Falasco}
\affiliation{Department of Physics and Astronomy,
University of Padova,
Via Marzolo 8, I-35131 Padova,
Italy}

\date{\today}
\begin{abstract}
For overdamped Langevin systems subjected to weak thermal noise and nonconservative forces, we establish a connection  between Freidlin-Wentzell large deviations theory and stochastic thermodynamics. First, we derive a series expansion of the quasipotential around the detailed-balance solution, i.e. the system's free energy, and identify the condition for the linear response regime to hold even far from equilibrium. Second, we prove that the escape rate from dissipative fixed points of the macroscopic dynamics is bounded by the entropy production of trajectories that relax into, and escape from the attractors. These results provide the foundation to study the nonequilibrium thermodynamics of dissipative metastable states.
\end{abstract}
\maketitle

\section{Introduction}

Metastability is a wide-spread phenomenon that can be observed in natural systems, spanning from climate science \cite{Margazoglou2021}, chemistry or biology \cite{assaf2013extrinsic}, as well as in man-made technological devices, like electronic bit-storage elements \cite{freitas2022reliability}. 
Heuristically, metastable states are only transiently stable, as they relax to the actual stable state by (e.g. thermal) fluctuations on exponentially long times.  

For systems that enjoy detailed-balance dynamics, i.e., that experience only conservative forces and thermal noise, metastable states are local free-energy minima separated by barriers (high enough with respect to the thermal energy) from their stable, thermodynamic equilibrium. Diamond and supercooled water are two standard examples, whose corresponding equilibrium (graphene and ice) can only be reached through nucleation.
The most suitable setup in which this phenomenon has been studied during the decades is the one of nonlinear systems described by overdamped Langevin dynamics, in the limit of weak-noise, corresponding to a sufficiently low temperature. 
Rarity of jumps among metastable states is then caused by the feeble effect of the thermal noise acting on the system.
Within this framework, the description of metastable states for detailed-balance systems was derived already in the early XX century by Eyring \cite{Eyring1935activated} and Kramers \cite{kramers1940brownian}, following the original ideas of Arrhenius \cite{Arrhenius1889}, to link the life-time of these states to the system energetics. In fact, for equilibrium systems where metastable states are (free) energy minima, the exit time $\tau$, i.e. the inverse of the escape rate $r$, depends exponentially on the barrier height $\Delta E$ enclosing a minimum, over the thermal energy $k_B T$,
\begin{equation}
    \tau_\text{eq} = r^{-1}_\text{eq} \asymp \mathrm{exp}\left(\dfrac{\Delta E}{k_BT}\right).
    \label{eq Arrhenius}
\end{equation}
This result holds in the weak noise limit, i.e. $\Delta E \gg k_BT$, and is today best understood within the framework of large deviations theory. The subexponential prefactor that sets the overall timescale is given by the Eyring-Kramers formula \cite{Haenggi1990kramers}.

For systems that are subject to nonconservative forces, thus continuously producing entropy, one can no longer rely on the free energy concept to describe metastable states through its minima and barrier. The most suitable definition of metastability is of pure dynamical nature: metastable Markovian systems are characterized by disparate characteristic times, as expressed by the spectrum of their stochastic generator, separated by diverging gaps.

For the case of dissipative systems under the action of a weak noise, Freidlin and Wentzell \cite{Freidlin1998} developed a theory in which the exponential exit time from nonequilibrium metastable states is reminiscent of Eq. \eqref{eq Arrhenius}, but the role of the energy is played by the \emph{quasipotential}  $I^\text{ss}$ \cite{Bouchet2014}:
\begin{equation}
    \tau= r^{-1}\asymp  \mathrm{exp}\left(\dfrac{\Delta I^\text{ss}}{\epsilon}\right). 
    \label{eq NoneqTau}
\end{equation}
The quasipotential is defined within large deviation theory \cite{bouchet2016generalisation}, when probabilities can be reasonably described according to a WKB ansatz in some limit of weak noise. For systems described by the Langevin dynamics of coordinates $x$ we have that $p(x,t)\asymp \mathrm{exp}(-I(x,t)/\epsilon)$, where $I(x,t)$ is the rate function associated to the system state $x$ and $\epsilon\to0^+$ is a small bookkeeping parameter that measures the noise intensity. Its stationary limit, $I^\text{ss}(x)= \lim_{t \to \infty} I(x,t)$, behaves like a potential function for the nonequilibrium dynamics, from which the name quasipotential \cite{Touchette2009}. Indeed, it is a Lyapunov function for the noiseless dynamics obtained by setting $\epsilon=0$ and determines the life-time of attractors through Eq. \eqref{eq NoneqTau}.
In general, the quasipotential $I^\text{ss}$ is not known a priori, reducing to the scaled system energy $E/{k_\text{B} T}$ only in the case of detailed-balance. For dissipative systems, $I^\text{ss}$ is hard to evaluate analytically by solving the small-noise expansion of the Fokker-Planck equation \cite{graham86,graham1987macroscopic} and requires dedicated numerical techniques to be extracted from simulations \cite{weinan2004minimum,grafke2017long}. With no access to the quasipotential, Eq. \eqref{eq NoneqTau} is far from being insightful. 

In this paper, we address the problem of metastability for thermodynamic systems following non-detailed-balance dynamics \cite{avanzini2024methods}, which can be described by overdamped diffusive processes. For this class of systems, ubiquitous in, e.g., soft matter physics, we provide a link between Freidlin-Wentzell theory and the nonequilibrium thermodynamics of relaxation within, and escape from isolated attractors. First, we derive an iterative expansion for the quasipotential in terms of the driving force breaking detailed-balance. This expansion can be readily implemented numerically, as it exploits the deterministic equilibrium relaxation only. 
Second, we show that the nonequilibrium escape rate $r$ is controlled by thermodynamic bounds. Knowing that transition rates are limited by measurable thermodynamic quantities can be useful to gain knowledge on system dynamics when the exact expression is not available. The upper bound on $\ln r$ is proven to be the dissipation along the most likely exit trajectory, called \emph{instanton}, while the lower bound is set by the negative of the dissipation along the relaxation trajectory. These two bounds, previously derived for Markov jump processes admitting a macroscopic limit \cite{freitas2022emergent,falasco2023macroscopic}, are here extended to diffusion processes using the representation of path probabilities in terms of physical coordinates only, and the orthogonal  decomposition of the drift field  \cite{Bertini2015Jun,zhou2016construction}. 
The bounds saturate for detailed-balance dynamics and close to it, where relaxation and escape trajectory are mapped on each other via time-reversal.

The paper is structured as follows: in Sec. \ref{sec Setup} the basic setup is provided, defining the stochastic process treated and setting the notation. In Sec. \ref{sec AvgThermo} we briefly recapitulate the stochastic thermodynamics \cite{Peliti2021,Seifert2012,VandenBroeck2015Jan,Gaspard2022,shiraishi2023introduction} of overdamped Langevin systems at the level of average quantities. In Sec. \ref{sec:WeakNoise} we derive the associated weak-noise limit by means of the large deviations approach. The emergent second law, previously obtained for Markov jump processes \cite{freitas2022emergent}, is shown to hold true for overdamped Langevin dynamics as well. In Sec. \ref{sec:noneq-exp} the nonequilibrium expansion for the quasipotential is given, showing that at first order (i.e. linear response regime) saturates the aforementioned emergent second law. Sec. \ref{sec Decomposition} introduces the  orthogonal decomposition of the drift vector field, showing also that the linear response approximation of the rate function holds in the nonlinear regime when some symmetry conditions are met. Sec. \ref{sec:bounds} is devoted to the derivation of the thermodynamic bounds on escape rates which are exemplified in Sec. \ref{sec:example} for a two-dimensional system consisting of a Brownian particle trapped in a double-well potential under the effect of a shear flow.

\section{Overdamped diffusion processes}
\label{sec Setup}

We consider a stochastic process described by a time-dependent probability distribution $P(\bm{x})$ over $\bm{x}\in \mathbb{R}^N$, whose evolution is given by a general Fokker-Planck equation,
\begin{equation}
d_t P(\bm{x}) = -\partial_{n} J_n(\bm{x}), 
\label{eq:FP}
\end{equation}
where $\partial_n \equiv \partial_{x_n}$, repeated indices are summed over, and the current $J_n$ is given by
\begin{equation}
J_n(\bm{x}) = \mu_n(\bm{x}) P(\bm{x}) - D_{nm}(\bm{x}) \partial_{m} P(\bm{x}).
\label{eq:FP_current}
\end{equation}
Here, $\mu_n(\bm{x})$ is a drift vector $D_{nm}(\bm{x})$ is a positive-definite diffusion matrix, which are assumed to be time-independent (i.e., we restrict to autonomous dynamics). The time dependence of $P(\bm{x})$ and $J_n(\bm{x})$ will not be written explicitly to ease notation. We assume there exists a unique steady state $P^\text{ss}(\bm{x})$,
with the corresponding current $J^\text{ss}_n(\bm{x})$ satisfying:
\begin{equation}
0 = \partial_n J_n^\text{ss}(\bm{x}).
\end{equation}
One special kind of steady states are equilibrium states, for which $J_n^\text{ss}(\bm{x})=0$ for all $n$ and $\bm{x}$. They exist if and only if the drift $\mu_n(\bm{x})$ derives from a gradient, in the sense that 
$\mu_n(\bm{x}) = -D_{nm}(\bm{x}) \partial_m \Phi(\bm{x})$ for some state function $\Phi(\bm{x})$. In that case, the steady-state distribution is $P^\text{eq}(\bm{x}) = e^{-\Phi(\bm{x})}/Z^\text{eq}$, with $Z^\text{eq} = \int d\bm{x} e^{-\Phi(\bm{x})}$. In thermodynamic systems that are interacting with a single thermal bath and are not subjected to nonequilibrium forces, the state function $\Phi(\bm{x})$ is just given by $\Phi(\bm{x}) = \beta E(\bm{x})$ in terms of the energy $E(\bm{x})$ of state $\bm{x}$ and the inverse temperature $\beta$. In the following we consider $k_B=\beta =1$ whenever we make reference to the thermodynamic interpretation. In general, we can always write the drift as
\begin{equation}
    \mu_n(\bm{x}) =  -D_{nm}(\bm{x}) \partial_m \Phi(\bm{x}) + f_n(\bm{x}) ,
   \label{eq:drift}
\end{equation}
where $\Phi(\bm{x})$ is an arbitrary state function and $f_n(\bm{x})$ is a force, which might or might not derive from the gradient of a state function. 
In the first case the force $f_n(\bm{x})$ 
will make the system reach a new equilibrium state. In the 
second case the system will reach a nonequilibrium steady state with 
persistent currents $J_n^\text{ss}(\bm{x}) \neq 0$.

We note that the Fokker-Planck equation can be alternatively written as 
\begin{equation}
    d_t \mathcal{I}(\bm{x}) = \partial_n j_n(\bm{x}) - \partial_n \mathcal{I}(\bm{x}) j_n(\bm{x})
    \label{eq:FP_selfinf}
\end{equation}
in terms of the self-information or surprisal 
\begin{equation}
    \mathcal{I}(\bm{x}) = -\ln (P(\bm{x})),
\end{equation}
and the reduced currents or probability velocities
\begin{equation}\label{eq:prob_velocity}
     j_n(\bm{x}) \equiv J_n(\bm{x})/P(\bm{x}) = \mu_n(\bm{x}) + D_{nm}(\bm{x}) \: \partial_m \mathcal{I}(\bm{x}).
\end{equation}

\section{Average Thermodynamics}
\label{sec AvgThermo}

Following the existing thermodynamic interpretation \cite{Peliti2021, Seifert2012}, we derive now the first and second law of thermodynamics in this setting, focusing on average quantities. 

\subsection{First Law}
We first consider the expectation value of the potential 
$\meann{\Phi} = \int d\bm{x} \: \Phi(x) P(\bm{x})$.
Its time derivative can be shown to be
\begin{equation}
   d_t \meann{\Phi} = \dot Q + \dot W
   \label{eq:first_law}
\end{equation}
in terms of
\begin{equation}
    \dot Q = - \int d\bm{x} \: j_n(\bm{x}) D^{-1}_{nm}(\bm{x}) \mu_m(\bm{x}) \: P(\bm{x}) \equiv - \mean{jD^{-1} \mu}
    \label{eq:heat_rate}
\end{equation}
and
\begin{equation}
    \dot W = \int d\bm{x} \: j_n(\bm{x}) D^{-1}_{nm}(\bm{x}) f_m(\bm{x}) \: P(\bm{x}) \equiv \mean{jD^{-1}f}
    \label{eq:work_rate}
\end{equation}
that are interpreted as the rate of heat (energy exchange with the environment) and work (energy provided to the system by the external force $f_n(\bm{x})$), respectively. This thermodynamic interpretation is a guide that will help us interpret the coming results, that apply to general diffusive processes even if they are not related to thermodynamics in any way. Thus, Eqs. \eqref{eq:heat_rate}, \eqref{eq:work_rate} can be considered definitions that, in combination with the general decomposition of the drift in Eq. \eqref{eq:drift}, make the energy balance in Eq. \eqref{eq:first_law} valid.

\subsection{Second Law}

We now consider the Shannon entropy $S = -\int d\bm{x} P(\bm{x})\ln (P(\bm{x}))$. Its time derivative can be written as
\begin{equation}
    d_t S = \mean{j D^{-1}j} -
    \mean{j D^{-1}\mu},
\end{equation}
from which we obtain the usual expression for the second law of thermodynamics
\begin{equation}
    \dot \Sigma = d_t S - \dot Q = \mean{j D^{-1}j} \geq 0.
    \label{eq:second_law}
\end{equation}
Since we are considering unit temperature, we can think of $-\dot Q$
as the rate of entropy change in the thermal environment, and therefore $\dot \Sigma$ is the mean total entropy production rate, which according to the previous expression is manifestly positive.

The second law can also be understood in terms of a nonequilibrium free energy, that is defined as the relative entropy between the instantaneous distribution and the equilibrium one:
\begin{equation}
\begin{split}
    \mathcal{F} &= \int d\bm{x} \: P(\bm{x})\ln (P(\bm{x})/P^\text{eq}(\bm{x})) \\
    & = \mean{\Phi} - S + \ln (Z^\text{eq}).
    \label{eq:free_energy}
\end{split}
\end{equation}
Then, Eq. \eqref{eq:second_law} can be rewritten as
\begin{equation}
\dot \Sigma = \dot W - d_t \mathcal{F} \geq 0.
\end{equation}

\subsection{Adiabatic/nonadiabatic decomposition}

In analogy with the nonequilibrium free energy in Eq. \eqref{eq:free_energy}, we can define the relative entropy between the instantaneous distribution and the steady state one
\begin{equation}
    \mathcal{G} \equiv \int d\bm{x} \: P(\bm{x}) \ln (P(\bm{x})/P^\text{ss}(\bm{x})),
\end{equation}
and computing its time derivative we find
\begin{equation}
    \dot \Sigma = \dot \Sigma_\text{a} + \dot \Sigma_\text{na},
\end{equation}
where
\begin{equation}
    \dot \Sigma_\text{a} \!=\! \meann{jD^{-1}j^\text{ss}}
    \quad , \quad 
    \dot \Sigma_\text{na} \!=\! -d_t \mathcal{G} \!=\! \meann{jD^{-1}(j-j^\text{ss})}
\end{equation}
are the adiabatic and nonadiabatic contributions to the entropy production. An important property of $\dot \Sigma_\text{a}$ and 
$\dot \Sigma_\text{na}$ is that they are always positive or zero. To see that, 
we first note that
\begin{equation}
\begin{split}
    \mean{j^\text{ss} D^{-1} (j-j^\text{ss})} &= 
    \int d\bm{x} \: J^\text{ss}_n(\bm{x}) \: \partial_n \ln \left(\!\frac{P^\text{ss}(\bm{x})}{P(\bm{x})}\!\right) \frac{P(\bm{x})}{P^\text{ss}(\bm{x})}\\
    & = -\int d\bm{x} \: J^\text{ss}_n(\bm{x}) \: \partial_n \left(\frac{P(\bm{x})}{P^\text{ss}(\bm{x})}\right)\\
    & = 0,
\end{split}
\end{equation}
where the last equality is obtained integrating by parts and using that $\partial_n J_n^\text{ss}(\bm{x}) = 0$. Then, it follows that $\dot \Sigma_\text{a}$ and $\dot \Sigma_\text{na}$ accept the following explicitly positive expressions:
\begin{equation}
    \dot \Sigma_\text{a} = \meann{j^\text{ss}D^{-1}j^\text{ss}}
\end{equation}
and 
\begin{equation}
    \dot \Sigma_\text{na} = -d_t \mathcal{G}= \meann{(j-j^\text{ss})D^{-1}(j-j^\text{ss})},
    \label{eq: NonadiabEPR}
\end{equation}
Note that $\dot \Sigma_\text{na} \geq 0$ implies that $\mathcal{G}$ is a Lyapunov function of the stochastic dynamics.

\section{Macroscopic or Weak-noise limit}
\label{sec:WeakNoise}

We consider now that the system has a scale parameter $\Omega$ and that $D_{nm}(\bm{x}) \propto 1/\Omega$ asymptotically in $\Omega \to \infty$. Then, the solution $P(\bm{x})$
to the Fokker-Planck equation satisfies a large deviations (LD) principle
\begin{equation}
    P(\bm{x}) \underset{{\Omega \to \infty}}{\asymp} e^{-\Omega I(\bm{x})},
\end{equation}
and the rate function $I(\bm{x})$ evolves according to
\begin{equation}
    d_t I(\bm{x}) = -\partial_n I(\bm{x}) \: u_n(\bm{x}) 
    - \partial_n I(\bm{x}) \: d_{nm}(\bm{x}) \: \partial_m I(\bm{x}),
    \label{eq:evol_rf}
\end{equation}
where $d_{nm}(\bm{x}) \equiv \lim_{\Omega \to \infty} \Omega D_{nm}(\bm{x})$
is the scaled diffusion matrix and 
\begin{equation}
    u_n(\bm{x}) \equiv  \lim_{\Omega \to \infty} \mu_n(\bm{x}) = 
    -d_{nm}(\bm{x}) \partial_m\phi(\bm{x}) + f_m(\bm{x}).
    \label{eq:scaled_drift}
\end{equation} 
The density $\phi(\bm{x})$ in the previous equation is defined in the following way: for any quantity $A$, we will consider its scaled version or density as $a \equiv \lim_{\Omega \to \infty} A/\Omega$, and the same will be done for the different thermodynamic quantities.
Equation \eqref{eq:evol_rf} is obtained by plugging the LD ansatz in Eq. \eqref{eq:FP_selfinf} and keeping the dominant terms in $\Omega$. Note that according to the LD principle, the rate function $I(\bm{x})$ can be interpreted as the self-information density $\mathcal{I}(\bm{x})/\Omega$ for $\Omega \to \infty$.
The most probable state at a given time is the global minimum $\bm{x}_t$ of the instantaneous rate function $I(\bm{x})$. It can be shown to evolve according to the closed deterministic dynamics
\begin{equation}
    d_t \bm{x}_t = \bm{u}(\bm{x}_t).
    \label{eq:det_dynamics}
\end{equation}

We now evaluate the first and second laws, as well as the adiabatic/nonadiabatic decomposition, in the limit $\Omega \to \infty$. Eq. \eqref{eq:first_law} reduces to:
\begin{equation}
    d_t \phi(\bm{x}_t) = \dot q(\bm{x}_t) + \dot w(\bm{x}_t),
\end{equation}
where 
\begin{equation}
    \dot q(\bm{x}) = -u_n(\bm{x})\: d^{-1}_{nm}(\bm{x}) \: u_m(\bm{x})
    \label{eq:heat_rate_det}
\end{equation}
and
\begin{equation}
    \dot w(\bm{x}) = u_n(\bm{x}) \: d^{-1}_{nm}(\bm{x}) \: f_m(\bm{x})
    \label{eq:work_rate_det}
\end{equation}
are the scaled heat and work rates, respectively. The second law in 
Eq. \eqref{eq:second_law} reduces to
\begin{equation}
    \dot \sigma(\bm{x}_t) = -\dot q(\bm{x}_t) \geq 0.
    \label{eq:second_law_det}
\end{equation}
Note that, $d_t s \to 0$ for $\Omega \to \infty$, which is natural since according to the LD principle $S$ scales as $\propto \ln \Omega$, so $s \to 0$ for $\Omega \to \infty$: there is no uncertainty on the microscopic state of the system in the absence of noise.

Evaluating the relative entropy $\mathcal{G}$ in the $\Omega \to \infty$ limit we obtain
\begin{equation}
   \lim_{\Omega \to \infty} \mathcal{G}/\Omega  = I^\text{ss}(\bm{x}_t),
\end{equation}
where $I^\text{ss}(\bm{x})$ is the steady-state rate function. Then, 
the scaled nonadiabatic entropy production rate, that is the scaled version of Eq. \eqref{eq: NonadiabEPR}, is $\dot \sigma_\text{na} = -d_t I^\text{ss}(\bm{x}_t)$, and the fact that it is positive says that the $I^\text{ss}(\bm{x})$ is a Lyapunov function of the deterministic dynamics. The positivity of the adiabatic entropy production rate implies that
\begin{equation}
    d_t I^\text{ss}(\bm{x}_t) + \dot \sigma(\bm{x}_t) \geq 0
    \label{eq:esl},
\end{equation}
which is the emergent second law identified in \cite{freitas2022} for Markov jump processes. Eq. \eqref{eq:esl} is useful since it allows to bound changes of the steady-state self-information in terms of the entropy produced along deterministic dynamics, as we exemplify in Section \ref{sec:example}. A different and early derivation of this result, that apparently went unnoticed at the time, was presented in \cite{Gaveau1998Nov}. 

\section{nonequilibrium expansion}
\label{sec:noneq-exp}

We will now consider the expansion of the steady-state rate function $I^\text{ss}(\bm{x})$ in different powers of the force $f_n(\bm{x})$. In the first place we split the drift $u_n(\bm{x})$ 
into zero and first order contributions:
\begin{equation}
    u_n(\bm{x}) =  \underbrace{-d_{nm}(\bm{x}) \partial_m \phi(\bm{x})}_{u^{(0)}_n(\bm{x})} + 
    \underbrace{f_n(\bm{x})}_{u_n^{(1)}(\bm{x})}.
   \label{eq:drift_split}
\end{equation}
We also split the steady-state rate function in a similar way:
\begin{equation}
    I^\text{ss}(\bm{x}) = I^{(0)}(\bm{x}) + I^{(1)}(\bm{x}) + I^{(2)}(\bm{x}) + \cdots 
    \label{eq:exp_rf}
\end{equation}
Then, expanding Eq. \eqref{eq:evol_rf} at steady state we find:
\begin{equation}
\begin{split}
    0 = &-\partial_n I^{(k)}(\bm{x}) \: u_n^{(0)}(\bm{x}) 
    -\partial_n I^{(k-1)}(\bm{x}) \: u_n^{(1)}(\bm{x}) \\
    &- \sum_{j=0}^{k} \partial_n I^{(j)}(\bm{x}) \: d_{nm}(\bm{x}) \:\partial_m I^{(k-j)}(\bm{x}).
\end{split}
\end{equation}
Recognizing that $I^{(0)}(\bm{x}) = \phi(\bm{x})$ and that 
therefore $d_{nm}(\bm{x}) \partial_m I^{(0)}(\bm{x}) = -u_n^{(0)}(\bm{x})$, 
we obtain:
\begin{equation}
\begin{split}
    u_n^{(0)}(\bm{x})\: \partial_n I^{(k)}(\bm{x}) 
    &= \partial_n I^{(k-1)}(\bm{x}) \: f_n(\bm{x}) \\
    &+ \sum_{j=1}^{k-1} \partial_n I^{(j)}(\bm{x}) \: d_{nm}(\bm{x}) \:\partial_m I^{(k-j)}(\bm{x}).
\end{split}
\label{eq:rf_pert_1}
\end{equation}
An important feature of this equation is that it allows to recursively solve for  $I^{(k)}(\bm{x})$. We also notice that $u_n^{(0)}(\bm{x})\: \partial_n I^{(k)}(\bm{x}) = d_t I^{(k)}\left(\bm{x}_t^{(0)}\right)$, where  $\bm{x}^{(0)}_t$
is a trajectory solving the zeroth-order deterministic dynamics 
$d_t \bm{x}^{(0)}_t = \bm{u}^{(0)}\left(\bm{x}^{(0)}_t\right)$. In particular, 
the first order component $I^{(1)}(\bm{x})$ satisfies:
\begin{equation}
\begin{split}\label{eq:I1}
    d_t I^{(1)}\left(\bm{x}_t^{(0)}\right) &= \partial_n \phi\left(\bm{x}^{(0)}_t\right) 
    \: f_n\left(\bm{x}^{(0)}_t\right) \\
    &= -u_m^{(0)}\left(\bm{x}^{(0)}_t\right) 
    \: d_{mn}^{-1}\left(\bm{x}^{(0)}_t\right) \: f_n\left(\bm{x}^{(0)}_t\right) \\
    &= -\dot w^{(1)}\left(\bm{x}^{(0)}_t\right),
\end{split}
\end{equation}
where in the last line we have used Eq. \eqref{eq:work_rate_det} (note that $\dot w^{(1)}$ is the lowest order contribution to $\dot w$). 
We now introduce the notation $a^{[k]} = \sum_{j=0}^k a^{(k)}$ for the partial reconstruction of any quantity $a$ up to order $k$. 
Using that $d_t I^{(0)} = d_t \phi \simeq \dot q^{[1]} + \dot w^{(1)}$ to first order, we see that
\begin{equation}
d_t I^{[1]}\left(\bm{x}_t^{(0)}\right) =  
\dot q^{[1]}\left(\bm{x}_t^{(0)}\right) = 
- \dot \sigma^{[1]}\left(\bm{x}_t^{(0)}\right).
\label{eq:full_I1}
\end{equation}
By comparing the last equation with Eq. \eqref{eq:esl}, we see that 
the emergent second law is saturated to first order in the force $f_n(\bm{x})$ along zeroth-order trajectories.
We notice that this expansion is conceptually different from a previous one proposed in \cite{Nardini2016} as it takes the detailed-balance dynamics as reference state. As a result of this choice, the different orders can be calculated recursively only using the relaxation dynamics with no need to consider optimal fluctuating trajectories (see Sec.\ref{sec:bounds}).

\section{Orthogonal decomposition} 
\label{sec Decomposition}

Now we explore the consequences of decomposing the drift $\mu_n(\bm{x})$ in terms of the gradient of the the steady-state self-information $\mathcal{I}^\text{ss}(\bm{x}) = -\ln (P^\text{ss}(\bm{x}))$. Then, from Eq. \eqref{eq:prob_velocity} we have
\begin{equation}
    \mu_n(\bm{x}) =  -D_{nm}(\bm{x}) \partial_m \mathcal{I}^\text{ss}(\bm{x}) + j^\mathrm{ss}_n(\bm{x}) 
   \label{eq:drift_g}
\end{equation}
where $j^\mathrm{ss}_n(\bm{x})$ is the probability velocity at steady state. By evaluating Eq. \eqref{eq:FP_selfinf} at steady state, we find that the probability velocity must satisfy the relation
\begin{equation}\label{eq:stationary_FPE}
0 = \partial_n j^\mathrm{ss}_n(\bm{x}) - \partial_n \mathcal{I}^\text{ss}(\bm{x})j^\mathrm{ss}_n(\bm{x}),
\end{equation}
at each order in $\Omega$. Namely, if we consider the expansions for the probability velocity 
\begin{equation}
j^\text{ss}_n(\bm{x})= v_n(\bm{x}) + h_n(\bm{x})/\Omega + O(1/\Omega^2),
\end{equation}
and for the self-information 
\begin{equation}
  \mathcal{I}(\bm{x}) = \Omega I(\bm{x}) + K(\bm{x}) + O(1/\Omega),
\end{equation}
we obtain from Eq. \eqref{eq:stationary_FPE}:
\begin{align}\label{eq:order_by_order}
&v_n(\bm{x}) \partial_n I^\text{ss}(\bm{x})= 0, \\
&\partial_n v_n(\bm{x}) - v_n(\bm{x}) \partial_n K(\bm{x}) - h_n(\bm{x}) \partial_n I(\bm{x})=0.\label{eq:ortho_ss}
\end{align}
The first equation tells one that the macroscopic  probability velocity $v_n$ is orthogonal to the level sets of $I^\text{ss}$, while the second is an equation to determine the subleading correction $K$ to the quasipotential (see Appendix \ref{app:subleading}).

To recap, we will employ in the following the decomposition 
\begin{align}
   u_n(\bm{x})= -d_{nm}(\bm{x}) \partial_m I^\text{ss}(\bm{x}) + v_n(\bm{x}),
   \label{eq:ortho_scaled_drift}
\end{align}
with $v_n(\bm{x})= \lim_{\Omega \to \infty} j^\text{ss}_n(\bm{x})$ and $v_n(\bm{x})\partial_n I^\text{ss}(\bm{x})=0$.
(Because of these properties we can also write $v_n(\bm{x})= a_{nm}(\bm{x}) \partial_m I^\text{ss} (\bm{x})$, with $a_{nm}$ an antisymmetric matrix.)
This decomposition implies that the macroscopic limit of the mean adiabatic entropy production rate is 
\begin{align}
\dot \sigma_\text{a}  := \lim_{\Omega \to \infty} \frac 1 \Omega  \dot \Sigma_\text{a} =v_n d^{-1}_{nm} v_m \geq 0.
\label{eq ScaledS_a}
\end{align}

Moreover, let us write $I^\text{ss}=\phi + \psi$ with $\psi$ not necessarily small and look for an equation for $\psi$. First, we note from Eq. \eqref{eq:ortho_scaled_drift} and Eq. \eqref{eq:scaled_drift} that 
$v_n = f_n + d_{nm} \partial_m \psi$. Then, we can rewrite Eq. \eqref{eq:ortho_ss} as 
\begin{align}
    0= \underbrace{f_n \partial_n \phi}_{-\dot w^{(1)}}  + \underbrace{f_n \partial_n \psi + d_{nm} \partial_n \psi \partial_m \psi}_{v_n \partial_n \psi} +  \underbrace{d_{nm} \partial_n \phi \partial_n \psi}_{-d_t \psi(\bm{x}_t^{(0)})}.
\end{align}
With the orthogonality condition $0=v_n \partial_n I^\text{ss}=v_n \partial_n \phi + v_n \partial_n \psi $ we arrive at 
\begin{align}
d_t \psi(\bm{x}_t^{(0)})= -\dot w^{(1)} - v_n \partial_n \phi
\end{align}
which says that the linear response approximation in Eq. \eqref{eq:I1} in general works well where $v_n \partial_n \phi$ is small, i.e. where the macroscopic velocity current lies on a level hypersurface of the potential $\phi$.

\section{Thermodynamics of rare trajectories} \label{sec:bounds}

We consider the overdamped Langevin equation corresponding to the Fokker-Planck equation \eqref{eq:FP},
\begin{equation}
     d_t x_n = \mu_n(\bm{x}) + \sqrt{2} B_{nm}(\bm{x})\,\eta_m(t),
     \label{eq OverdampedLangevin}
\end{equation}
where the noise is Gaussian, zero-mean and white, namely 
\begin{equation}
    \langle\eta_n(t) \rangle=0 \quad , \quad \langle\eta_n(t)\eta_m(t')\rangle=\delta_{nm}\delta(t-t')
    \label{eq GaussianNoise}
\end{equation} 
and $B_{nk}(\bm{x})B_{km}(\bm{x}) = D_{nm}(\bm{x}) $. Since the noise in Eq. \eqref{eq OverdampedLangevin} is multiplicative one should choose the anti-Ito prescription to obtain the Fokker-Planck equation in Eq. \eqref{eq:FP}. However, all prescriptions are equivalent when we consider the leading order of the weak-noise limit. 

The asymptotic limit $\Omega\to\infty$ reduces Eq. \eqref{eq OverdampedLangevin} to the deterministic expression already introduced in Eq. \eqref{eq:det_dynamics}.
We define the fixed points $\{\bm{x^*}\}$ of the deterministic dynamics as the solution of the equations
\begin{equation}
   u_n(\bm{x^*}) = 0.
    \label{eq FixedPoints}
\end{equation}
Their stability can be inferred by studying the linearized dynamics in their vicinity \cite{Strogatz2018}:
\begin{equation}
d_t \delta x_n =\delta x_m \partial_m u_n(\bm{x^*}).
\end{equation}
If all eigenvalues of the Jacobian matrix with elements $\partial_m u_n(\bm{x^*})$ have the same (positive or negative) sign, $\bm{x^*}$ is called a (unstable or stable, resp.) node. Otherwise, it is called saddle. 
If there are two stable fixed points $\bm{x^*}_{(i)}, \bm{x^*}_{(j)}$ separated by a saddle $\bm{x^*}_{(\nu)}$, we say that the system is bistable. When a small noise is added the system displays metastability \cite{Arrhenius1889}. Note that nondetailed-balance dynamics admits more general attractor than fixed points, such as limit cycles, which will not be considered in the following.

The metastable state is defined as the basin of attraction of a stable fixed point, namely the set of points that if taken as initial condition for the noiseless dynamics of Eq. \eqref{eq:det_dynamics} would relax to the chosen fixed point \cite{Falasco2021}. 
A transition between states, being in the framework of weak noise, corresponds to an escape trajectory that starting in the fixed point $\bm{x^*}_{(i)}$ after an ideally infinite time reaches the saddle point $\bm{x^*}_{(\nu)}$ at the boundary with the basin of attraction of $\bm{x^*}_{(j)}$, i.e.
\begin{equation}
    \bm{x}(t=0) = \bm{x^*}_{(i)} \hspace{3mm},\hspace{3mm} \bm{x}(t\to\infty) = \bm{x^*}_{(\nu)}.
    \label{eq BoundaryInst}
\end{equation}
Once on the saddle, the trajectory follows a deterministic relaxation towards the target fixed point $\bm{x^*}_{(j)}$.

In the next paragraph we show how the problem of calculating transition rates between metastable states in systems lacking detailed-balance can be traced back to the search of most likely escape trajectories. The knowledge of this trajectory will lead to the derivation of a thermodynamic upper bound on the nonequilibrium transition rate.

\subsection{Optimal escape trajectory: the \emph{instanton}}




Considering the weak noise limit, the dynamics for the rate function in Eq. \eqref{eq:evol_rf} can be rewritten as a Hamilton-Jacobi equation:
\begin{equation}
    -\partial_t I(x) = H(\bm{x}, \bm{\pi} = \nabla I(x)),
    \label{eq:HJ}
\end{equation}
in terms of the Hamiltonian
\begin{equation}
    H(\bm{x}, \bm{\pi}) = \bm{\pi}\cdot \bm{u}(\bm{x}) + 
    \bm{\pi}^T \cdot \bm{d}(\bm{x}) \cdot \bm{\pi}
\end{equation}
where $\bm{d}(\bm{x})$ is the scaled diffusion matrix. In Eq. \eqref{eq:HJ}, the rate function acts as the action function and $\bm{\pi} = \nabla I(x)$ as the conjugate momentum.

The equations of motion corresponding to such Hamiltonian are:
\begin{align}
     d_t x_n &= u_n(\bm{x}) + 2 \: d_{nm}(\bm{x})\pi_m(\bm{x}) \label{eq HamiltonianSystem}\\
     d_t \pi_n &= -\pi_k \: \partial_n u_k(\bm{x}) - \pi_j \: \partial_n d_{jk}(\bm{x}) \: \pi_k. 
    \label{eq HamiltonianSystem2}
\end{align}
These equations can be alternatively derived from a different perspective, if the system is studied via the path integral approach. In fact, one can investigate the stochastic dynamics by looking at the path probability $P[\bm{x}]$ associated to the ensemble of trajectories $\{\bm{x}(t)\}^{\tau}_0$ constrained to the initial $\bm{x}(0)$ and final $\bm{x}(t)$, which is a generalization of the original Onsager-Machlup path probability to nonlinear systems \cite{Cugliandolo2017}. In the asymptotic limit, we write
\begin{align}
    P[\bm{x}] 
    &\asymp \mathrm{exp}\left( -\Omega \int_0^t d\tau\, \dfrac{(\dot{\bm{x}} - \bm{u}(\bm{x}))\bm{d}(\bm{x})^{-1}(\dot{\bm{x}} - \bm{u}(\bm{x}))}{4}\right) \nonumber \\
    &\equiv \mathrm{exp}(-\Omega\mathcal{A}[\bm{x}, \dot{\bm{x}}]).
    \label{eq pathProb}
\end{align}
A Hubbard-Stratonovich transformation allows to obtain the Hamiltonian form:
\begin{equation}
    P[\bm{x}] \asymp \int \mathcal{D}\bm{\pi}e^{-\Omega(\dot{\bm{x}}\cdot\bm{\pi} - H(\bm{x}, \bm{\pi}))}.
    \label{eq ActionHamilton}
\end{equation}
Then, Eq. \eqref{eq HamiltonianSystem}
is obtained 
by functional minimization of the action $\mathcal{A}$.

The system admits two classes of solutions that maximize the path probability (Fig. \ref{fig: InstAndRel}): solutions on the manifold $\bm{\pi}=0$, corresponding to relaxation trajectories, that in fact reduce Eq. \eqref{eq HamiltonianSystem} to $\dot{\bm{x}} = \bm{u}(\bm{x})$; solutions with $\bm{\pi}\neq 0$ corresponding to the most typical path of large fluctuations, called instantons.

In particular, we observe that when stationarity is reached for $t\to\infty$ in Eq. \eqref{eq:HJ}, we are left with $H(\bm{x}, \nabla I^\text{ss}(\bm{x}))=0$, which allows us to identify $\bm{\pi} = \nabla I^\text{ss}$. Namely, the instanton solution has to be found in the $H=0$ manifold. The result is the infinite-time trajectory that at the boundaries satisfies Eq. \eqref{eq BoundaryInst}.

At this point, the interest is on the derivation of the explicit equation of motion of the instanton. 
Replacing $\bm{\pi} = \nabla I^\text{ss}$
in Eq. \eqref{eq HamiltonianSystem}
we obtain:
\begin{equation}\label{eq:instanton_dyn}
\begin{split}
d_t x_n &= u_n(\bm{x}) + 2 \: d_{nm}(\bm{x})\partial_m I^\text{ss}(\bm{x})  \\
&= v_n(\bm{x}) + d_{nm}(\bm{x})\partial_m I^\text{ss}(\bm{x}),
\end{split}
\end{equation}
where in the second line we have employed the orthogonal decomposition of Eq. \eqref{eq:ortho_scaled_drift}.
Notice that for the relaxation dynamics, we would have the opposite sign for the gradient of the quasipotential, while the velocity term would be the same, as it is due to the current term which characterizes the stationary state. Then,
\begin{equation}
    d_t x_n = v_n(\bm{x}) \pm d_{nm}(\bm{x}) \partial_m I^\text{ss}(\bm{x}) \hspace{2mm} 
        \begin{cases}
            + \text{  if instanton}\\
            - \text{  if relaxation}.
        \end{cases}
        \label{eq CFRInstVSRel}
\end{equation}
The previous equation shows how the presence of currents in a nondetailed-balance system breaks the symmetry between the relaxation and the escape dynamics, which is retrieved for the case of zero nonconservative force $f=0$, where the instanton becomes the time-reversed relaxation trajectory.

\subsection{Bounding the transition rate}

It is known that the exit time from a metastable state obeys an exponential law, governed by a constant rate in the context of weak-noise \cite{Day1983,bouchet2016generalisation}. The rate can be obtained by using the condition $H=0$ in Eq. \eqref{eq ActionHamilton}, in the infinite time limit for the instanton. Indeed, we have:
\begin{equation}
\begin{split}
    \lim_{t\to\infty}&P[\bm{x}|\bm{x}(0)=\bm{x^*}_{(i)},\bm{x}(t)=\bm{x^*}_{(\nu)}] \\ &\asymp \mathrm{exp}\left(-\Omega\int d\tau\: \bm{\dot{x}}\cdot\nabla I^\text{ss}(\bm{x}(t))\bigg\rvert_{\bm{x^*}_{(i)}}^{\bm{x^*}_{(\nu)}}\right)\\
    &= e^{-\Omega(I^\text{ss}(\bm{x^*}_{(\nu)}) - I^\text{ss}(\bm{x^*}_{(i)}))}
\end{split}
\end{equation}
which allows one to write
\begin{equation}
    r_{\nu} \asymp e^{-\Omega(I^\text{ss}(\bm{x^*}_{(\nu)}) - I^\text{ss}(\bm{x^*}_{(i)}))},
    \label{eq NoneqRate}
\end{equation}
in full analogy to Arrhenius formula for detailed-balance dynamics, where the quasipotential is just the equilibrium potential, i.e. $I^\text{ss}=\Phi$. Equation \eqref{eq NoneqRate} can be shown to be controlled by a lower and an upper bound of thermodynamic nature.

To derive such thermodynamic bounds, the starting point is the general splitting of the entropy production for a stochastic process into its adiabatic and nonadiabatic terms \cite{Peliti2021}, analogously to what we did in section \ref{sec AvgThermo}. The difference in this case is that the following decomposition holds for fluctuating quantities
\begin{equation}
    \dot{\sigma} = \dot{\sigma}_{\text{a}} + \dot{\sigma}_{\text{na}}.
\end{equation}
These two contributions can be identified, along any trajectory, from the definition of entropy flow as the log-ratio of forward  and backward path probabilities, respectively  denoted $P[\bm{x}]$ and $P[\overline{\bm{x}}]$ \cite{falasco2023macroscopic}
\begin{equation}
    \sigma =\dfrac{1}{\Omega} \ln \dfrac{P[\bm{x}]}{P[\overline{\bm{x}}]}.
\end{equation}
Using Eq. \eqref{eq pathProb}, we obtain the explicit expression for the entropy flow along a trajectory of duration $t$,
\begin{equation}
    \sigma = \int_{0}^td\tau  \,\dot{\sigma}=  \int_{0}^td\tau  \,u_n(\bm{x})\,d_{nm}^{-1}(\bm{x})\,\dot{x}_m .
\end{equation}
Recalling the decomposition of the scaled drift term in Eq. \eqref{eq:ortho_scaled_drift}, valid in the  macroscopic limit $\Omega\to\infty$, we write
\begin{equation}
    \dot{\sigma} = (-d_{nm}(\bm{x})\partial_m I^\text{ss}(\bm{x}) + v_n)\,d_{nm}^{-1}(\bm{x})\,\dot{x}_m,
\end{equation}
and expanding this product, we naturally identify two terms
\begin{equation}
    \dot{\sigma} = -\partial_mI^\text{ss}(\bm{x})\,\dot{x}_m + v_n(\bm{x})\,d_{nm}^{-1}(\bm{x})\,\dot{x}_m.
    \label{eq EPRsplit}
\end{equation}
The first term in Eq. \eqref{eq EPRsplit},
\begin{equation}
    -\partial_mI^\text{ss}(\bm{x})\,\dot{x}_m = -d_tI^\text{ss}(\bm{x}) = \dot{\sigma}_{\text{na}}
\end{equation}
is the nonadiabatic entropy production rate, holding for any stochastic trajectory in the asymptotic limit. It reduces to the nonadiabatic contribution used in the emergent second law in Eq.\eqref{eq:esl}, once it is evaluated along the deterministic relaxation trajectory. For the second contribution in Eq. \eqref{eq EPRsplit}, we focus directly on the instanton and relaxation trajectories, $\dot x_m = \pm d_{nm}(\bm{x})\,\partial_m I^\text{ss}(\bm{x}) + v_n(\bm{x}) $, by writing
\begin{align}\label{eq_adiabaticEPR_traj}
  & 0 \leq  v_n(\bm{x})\,d_{nm}^{-1}(\bm{x})\,\dot{x}_m 
    = v_n(\bm{x})\,d_{nm}^{-1}(\bm{x})v_n(\bm{x}) = \dot{\sigma}_{\text{a}},
\end{align}
thanks to the orthogonality condition in Eq. \eqref{eq:order_by_order}. The adiabatic entropy production rate given by Eq. \eqref{eq_adiabaticEPR_traj} extends the result valid for the relaxation dynamics of Eq. \eqref{eq ScaledS_a} to the instanton trajectory.

The fact that Eq. \eqref{eq_adiabaticEPR_traj} is non-negative irrespective of the dynamics considered, be it relaxational or instantonic, allows one to derive thermodynamics bound on the transition rate of Eq. \eqref{eq NoneqRate}. Indeed, the nonadiabatic entropy production of the relaxation $\sigma_{\text{na}}^{\nu\to i}$ is equal and opposite to the the nonadiabatic entropy production of the instanton $\sigma_{\text{na}}^{i\to \nu}$, being the difference between initial and final values of the quasi potential, 
\begin{equation}
    \sigma_{\text{na}}^{\nu\to i} = \Delta I^\text{ss} =  -  \sigma_{\text{na}}^{i\to\nu} .
\end{equation}
Hereafter, $i\to\nu$ and $\nu\to i$ will denote the instanton and the relaxation trajectory, respectively, connecting the saddle $\nu$ with the stable fixed point $i$.

As a consequence, the following two inequalities hold
\begin{align}
    \sigma_{i\to\nu} &= \sigma_{\text{a}} - \Delta I^\text{ss} \geq - \Delta I^\text{ss},\\
    \sigma_{\nu\to i} &= \sigma_{\text{a}} + \Delta I^\text{ss} \geq  \Delta I^\text{ss},
\end{align}
which are equivalent to
\begin{align}
    e^{\Omega\, \sigma_{i \to \nu}} &\geq e^{-\Omega\,\Delta I^\text{ss}},\\
   e^{-\Omega \,\sigma_{\nu \to i}} &\leq e^{-\Omega\,\Delta I^\text{ss}}.
\end{align}
The right-hand side of both inequalities corresponds to the leading order of the nonequilibrium transition rate appearing in Eq. \eqref{eq NoneqRate}, yielding the thermodynamic bounds on the transition rate 
\begin{equation}
    -\sigma_{\nu\to i} \leq \lim_{\Omega \to \infty} \frac{1}{\Omega}\ln r_{\nu} \leq \sigma_{i\to\nu}.
    \label{eq BOUNDS}
\end{equation}

{\color{black}

\section{Example}
\label{sec:example}

    In this section, we illustrate our previous results by looking at the specific example of the two-dimensional double-well subjected to a shear flow, which breaks detailed-balance. 
    First, in Sec. \ref{subsec:model} we specify the model characteristics and study its deterministic dynamics. 
    Then, in Sec. \ref{subsec: ESL} we exemplify the use of the emergent second law, exploiting the fact that we can use entropy production as an upper bound for the steady state rate function $I^{\text{ss}}$. To do so, we rely on an empirical estimate of such function, $I^{\text{ss}}_{\text{est}}$ . This estimate is obtained by deriving a histogram of the steady state distribution along a reaction coordinate, as we will explain in the second paragraph of this section.
    After that, in Sec. \ref{subsec: noneqExpansion} we make use of the nonequilibrium expansion of the quasipotential in Eq. \eqref{eq:rf_pert_1}, limiting ourselves to first order in the shear, and show there is good agreement when comparing with the aforementioned $I^{\text{ss}}_{\text{est}}$.
    Finally, in Sec. \ref{subsec: ThDynBounds} we test the thermodynamic bounds of Eq. \eqref{eq BOUNDS}, after dealing with the problem of obtaining the instanton trajectory, which requires a sophisticated numerical technique.

\subsection{The double-well with shear flow}\label{subsec:model}


Our two-dimensional model is described by the following Langevin dynamics
\begin{equation}
\begin{split}
    d_t x_1 &= \mu_1(\bm{x}) + \sqrt{2/\Omega} \: \eta_1(t), \\
    d_t x_2 &= \mu_2(\bm{x}) + \sqrt{2/\Omega} \: \eta_2(t) ,
\end{split}
\label{eq:example_SDE}
\end{equation}
where $\eta_i(t)$ are zero-mean Gaussian variables satisfying Eq. \eqref{eq GaussianNoise}. This corresponds to taking $D_{nm} (\bm{x}) = \Omega^{-1} \mathbb{I}_{2\times 2}$ in Eq. \eqref{eq:FP_current}. For the drift we consider the combination of a conservative force coming from a bistable potential and a non-conservative rotational field,
\begin{equation}
    \mu_n(\bm{x}) = -\Omega^{-1} \partial_n\Phi(\bm{x}) + f_n(\bm{x}),
\end{equation}
where the potential $\Phi(\bm{x})$ is given by 
\begin{equation}
    \Phi(\bm{x}) = \Omega(\alpha x_1^4 - \beta x_1^2 + x_2^2),
\end{equation}
and the force $f_n(\bm{x})$, corresponding to a shear flow, reads
\begin{equation}
    f_1(\bm{x}) = -\gamma x_2 \qquad f_2(\bm{x}) = \gamma x_1.
\end{equation}
The constants $\alpha$, $\beta$ and $\gamma$ are free parameters. 
Figure \ref{fig:potential_force} shows the equipotential lines of 
$\Phi(\bm{x})/\Omega$ and the direction of the force $\bm{f}(\bm{x})$. In Figure \ref{fig:hists} we show the steady state distributions for different strengths of the force  $\bm{f}(\bm{x})$, that were obtained from stochastic trajectories generated by the direct numerical integration of Eqs. \eqref{eq:example_SDE}.

\begin{figure}
    \centering
    \includegraphics[scale=.7]{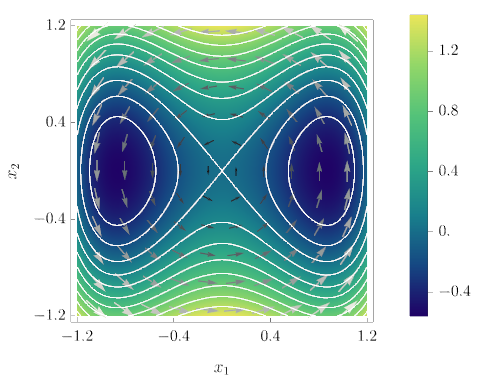}
    \caption{Equipotential lines of $\Phi(\bm{x})/\Omega$. The arrows indicate the direction of the rotational field $\bm{f}(\bm{x})$. The parameters are $\alpha=1$, $\beta=1.5$ and $\gamma=1$.}
    \label{fig:potential_force}
\end{figure}

\begin{figure}[ht!]
    \centering
    \includegraphics[scale=.5]{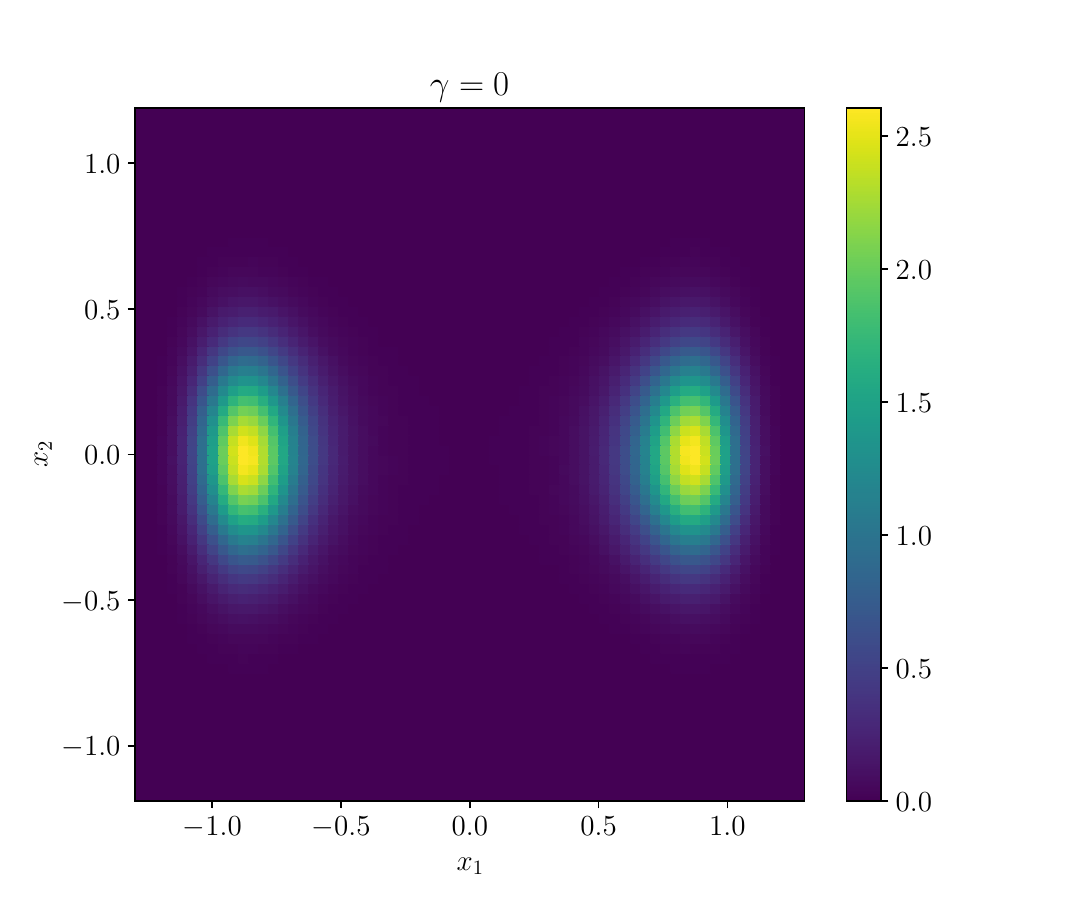}
    \includegraphics[scale=.5]{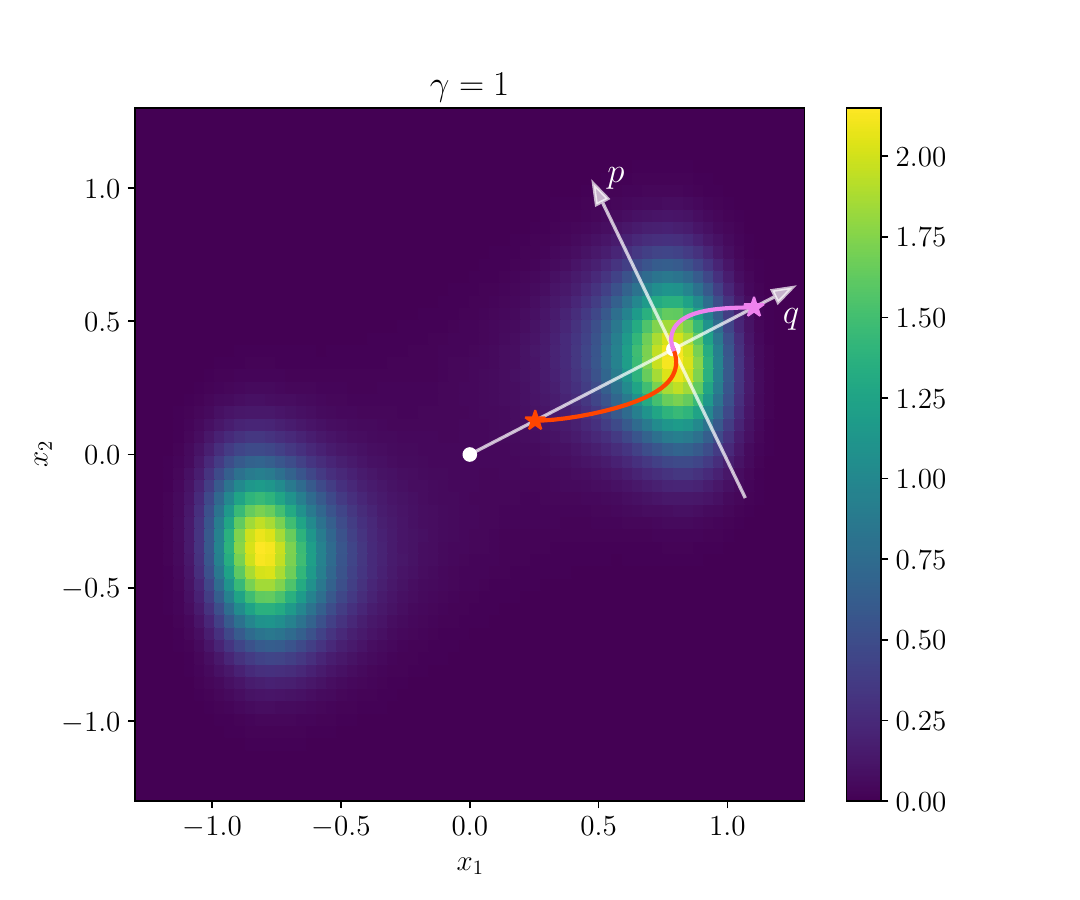}
    \caption{Steady state density histograms for different values of the force parameter $\gamma$. The top panel corresponds to $\gamma=0$ and therefore shows the equilibrium distribution, while the lower panel is for $\gamma=1$. In the two cases we have $\alpha=1$, $\beta=1.5$ and $\Omega=10$. The data was obtained by numerical integration of Eqs. \eqref{eq:example_SDE}. In the lower panel we also show a new coordinate $q$ (joining the origin and one deterministic fixed point $\bm{x}^*$) and two deterministic trajectories starting at different points (indicated by stars) and relaxing towards $\bm{x}^*$.}
    \label{fig:hists}
\end{figure}

Before dealing with the emergent second law, it is useful to study the model deterministic dynamics $d_t \bm{x} = \bm{u}(\bm{x})$. It has three fixed points, $\bm{x}^* =0$ or 
\begin{equation}
\begin{split}
x^*_1 &= \pm \frac{1}{2} \sqrt{(4\beta - \gamma^2)/2\alpha},\\
x^*_2 &= \gamma \: x^*_1/2.
\end{split}
\end{equation}
A linear stability analysis shows that $\bm{x}^*=0$ is a stable fixed point only if $\gamma^2 > 4 \beta$, while the other two fixed points become stable otherwise. An estimation based on a Gaussian approximation shows that the dominant effect of noise in the deterministic equations of motion (tracking the evolution of the mean values $\mean{x_1}$) is to renormalize the constant $\beta$ as $\beta - 6\alpha \sigma_1^2$, where $\sigma_1^2$ is the variance of $x_1$, which decreases as $1/\Omega$ for large $\Omega$.

\subsection{Emergent second law}\label{subsec: ESL}

According to the emergent second law in Eq. \eqref{eq:esl}, given a trajectory $\bm{x}_t$ starting at a point $\bm{x}_0$, the macroscopic entropy production $\sigma(\bm{x}_0) = \int_0^{+\infty} d\tau \dot \sigma(\bm{x}_\tau)$ is an upper bound to $I^\text{ss}(\bm{x_0})$, the steady-state rate function evaluated at the initial state of this nonequilibrium relaxation trajectory. This is useful since the steady-state rate function is usually hard to obtain, while $\sigma(\bm{x}_0)$ can be computed from the deterministic dynamics alone. 
We implement this idea for points along the axis $q$ joining the origin $x_1=x_2=0$ with one deterministic fixed point $\bm{x}^*$. An example is shown in Figure \ref{fig:hists}. In first place, in Figure \ref{fig:hist_q} we show the histogram of the steady state distribution for different values of $q$ and $|p|<\epsilon=10^{-2}$, where $p$ is the coordinate orthogonal to $q$. From such histogram, it is possible to obtain an estimate of the rate function as $I_\text{est}^\text{ss}(q) = -\ln (P(q \:|\: |p|<\epsilon))/\Omega$. Of course, for $I_\text{est}^\text{ss}(q)$ to be an accurate estimate of the true rate function, the value of $\Omega$ must be high enough, which makes the direct sampling of large fluctuations increasingly difficult. In Figure \ref{fig:esl} 
we compare the estimate $I_\text{est}^\text{ss}(q_0)$ 
(obtained from data generated with $\Omega=30$), where $q_0$ is the starting point along reaction coordinate $q$, with the upper bound $\sigma(q_0)$. For $\gamma=0$ we know that the upper bound and the actual rate function coincide, since the dynamics is detailed-balance and the emergent second law is saturated in that case (indeed, the relaxation dynamics is actually the equilibrium one).
We see in Figure \ref{fig:esl} that this is actually the case, which indicates that $\Omega=30$ is already high enough to estimate the rate function. For $\gamma=1$ we see that $\sigma(q_0)$ indeed works as an upper bound to $I_\text{est}^\text{ss}(q_0)$. The fact that the bound is tighter to the right of the fixed point can be traced back to the increased relaxation speed of the corresponding trajectories (see \cite{freitas2022emergent} for a detailed analysis of a similar observation).

\begin{figure}
    \centering
    \includegraphics[scale=.56]{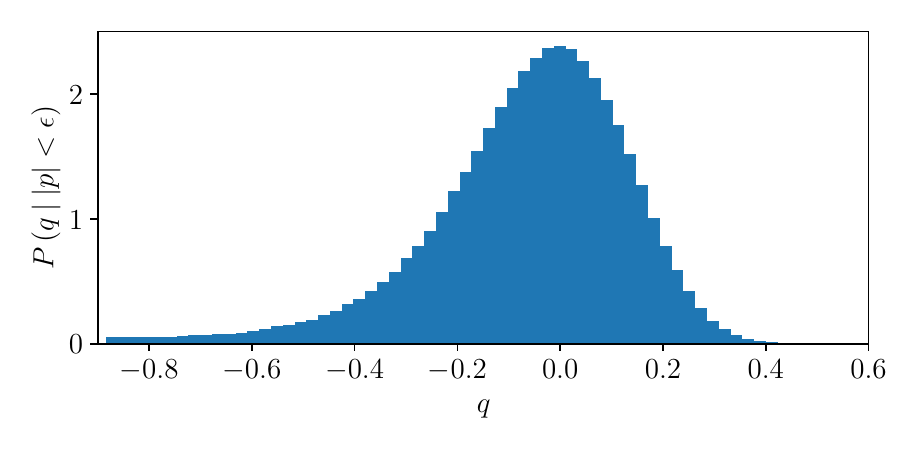}
    \caption{Density histogram along the axis $q$ in the lower panel of Figure \ref{fig:hists} ($\alpha=1$, $\beta=1.5$, $\gamma=1$, $\Omega=10$).}
    \label{fig:hist_q}
\end{figure}

\begin{figure}
    \centering
    \includegraphics[scale=.56]{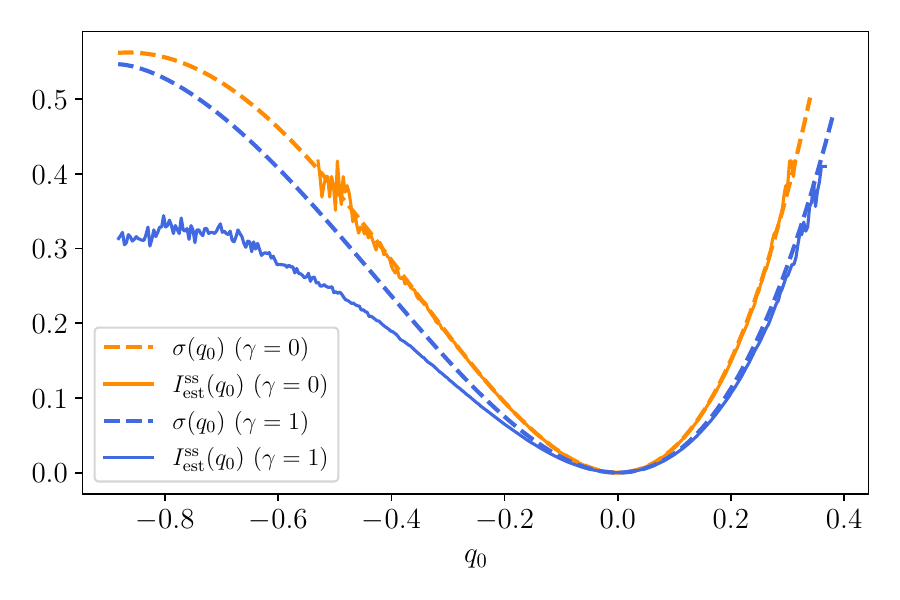}
    \caption{Comparison between the estimation of the rate function $I_\text{est}^\text{ss}(q_0)$ and the integrated entropy production along deterministic trajectories $\sigma(q_0)$ as a function of $q_0$, the starting point along axis $q$. The parameters are $\alpha=1$, $\beta=1.5$, and $\Omega=30$.}
    \label{fig:esl}
\end{figure}

\subsection{Nonequilibrium expansion}\label{subsec: noneqExpansion}

\begin{figure}[ht!]
     \centering
     \includegraphics[scale=.5]{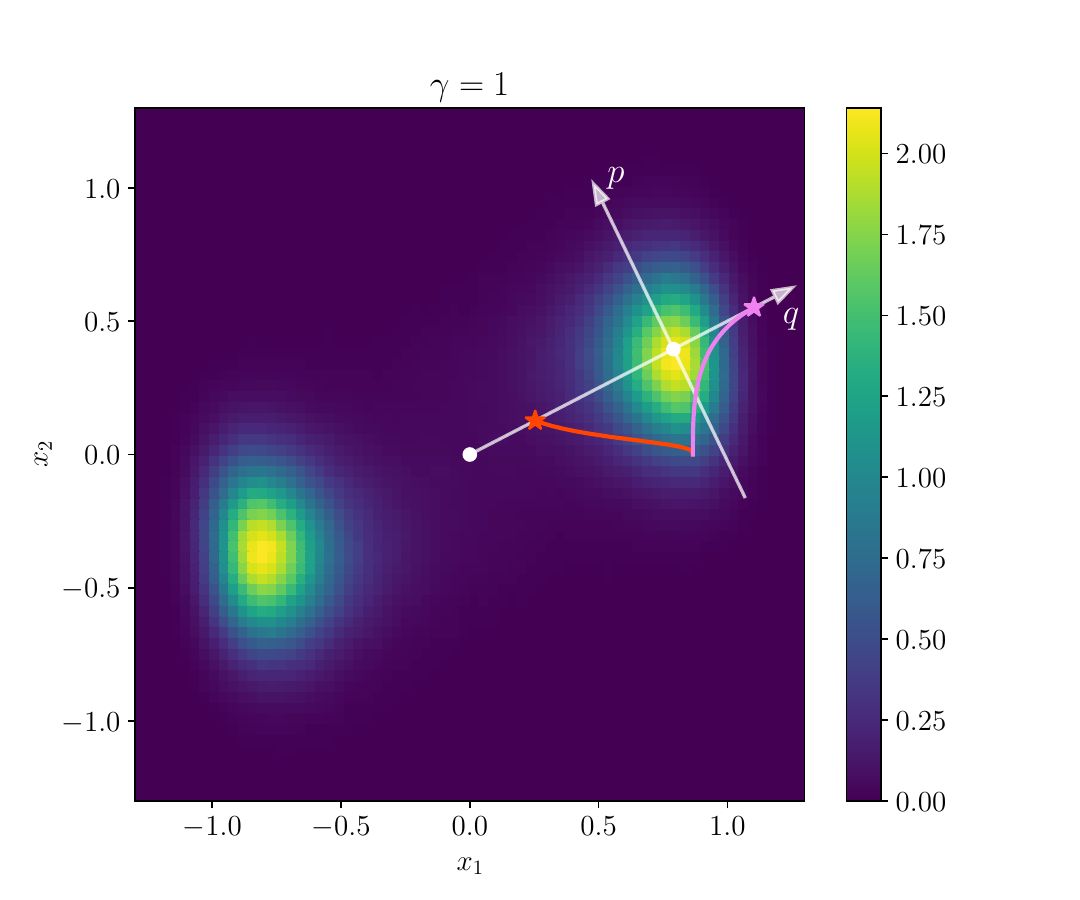}
     \caption{Steady state density histogram
     and two detailed-balanced deterministic trajectories starting at different points (indicated by stars) and relaxing towards $\bm{x}^{*(0)}$ (the fixed point of $d_t \bm{x}_t = \bm{u}^{(0)}\left(\bm{x}_t\right)$),
     for parameters $\alpha=1$, $\beta=1.5$, $\gamma=1$ and $\Omega=10$.} 
     \label{fig:hist_exp}
 \end{figure}

 \begin{figure}[ht!]
     \centering
     \includegraphics[scale=.55]{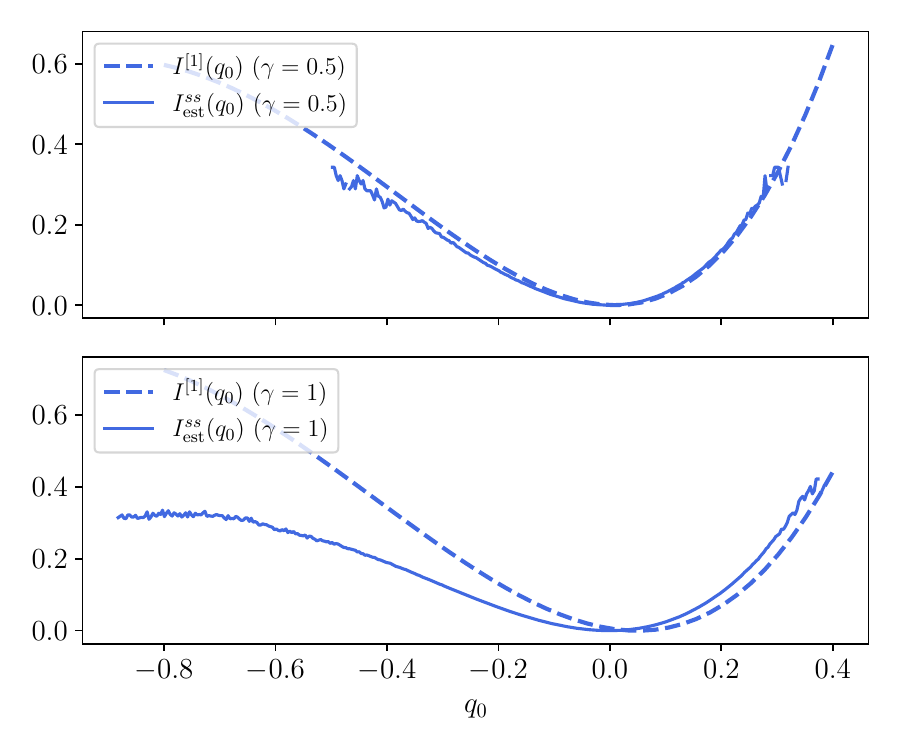}
     \caption{Comparison between the estimation of the rate function $I_\text{est}^\text{ss}(q_0)$ and the first order approximation $I^{[1]}(q_0)$ obtained from Eq. \eqref{eq:full_I1}, for two different values of $\gamma$ ($\alpha=1$, $\beta=1.5$).} 
     \label{fig:I_first_order}
 \end{figure}

 We now illustrate the use of the nonequilibrium expansion of Section \ref{sec:noneq-exp}. According to Eq. \eqref{eq:rf_pert_1}, contributions of different orders to the steady-sate rate function can be reconstructed by integrating different quantities over detailed-balanced deterministic trajectories, i.e., those satisfying $d_t \bm{x}^{(0)}_t = \bm{u}^{(0)}\left(\bm{x}^{(0)}_t\right)$. Examples of such trajectories are shown in Figure \ref{fig:hist_exp}. Note that they converge to the fixed point $\bm{x}^{*(0)}$ corresponding to $\gamma=0$, in contrast to the deterministic trajectories involved in the evaluation of the emergent second law, that converge to the true fixed point for given parameters (Figure \ref{fig:hists}).

 In Figure \ref{fig:I_first_order} we compare $I_\text{est}^\text{ss}(q_0)$, the numerical estimation of the rate function, with the first order approximation $I^{[1]}(q_0)$ obtained by integrating Eq. \eqref{eq:full_I1}. We see that the difference increases with $\gamma$, the strength of the rotational force field.

 In principle, it is possible to compute the second order correction to the rate function, which from Eq. \eqref{eq:rf_pert_1}
 satisfies:
 \begin{equation}
 \begin{split}
     u_n^{(0)}(\bm{x})\:\partial_n I^{(2)}(\bm{x})
     &= \partial_n I^{(1)}(\bm{x})
     \: f_n(\bm{x})\\
     &+ \partial_nI^{(1)}(\bm{x}) \: d_{nm}(\bm{x}) \: \partial_m I^{(1)}(\bm{x}).
 \end{split}
 \end{equation}
 Therefore, to obtain $I^{(2)}(\bm{x})$ by integration of the previous equation along detailed-balance deterministic trajectories, we need to first compute the gradient of $I^{(1)}(\bm{x})$ at each point. Eq. \eqref{eq:rf_pert_1} gives directly the scalar product of such gradient and the velocity $\bm{u}^{(0)}(\bm{x})$, but we also need the other components. In the two-dimensional example we are considering, there is only one additional component. Thus, a possible strategy to obtain the full gradient of $I^{(1)}(\bm{x})$ is to compute along two sufficiently close trajectories. It remains to be explored how to efficiently implement this procedure in high dimensional systems.

\subsection{Thermodynamic bounds}\label{subsec: ThDynBounds}

\begin{figure}[t]
	\centering 
	\includegraphics[width=0.45\textwidth, angle=0]{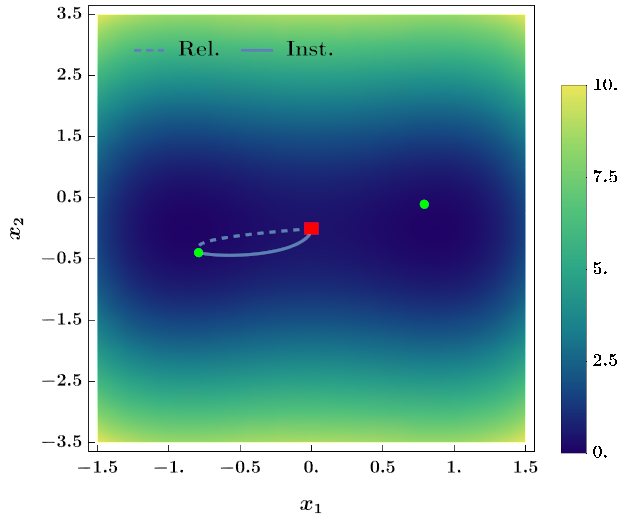}	
	\caption{\small Example of instanton (solid) and relaxation (dashed) trajectories, for parameters $\alpha=1$, $\beta=1.5$, $\gamma=1$. Green points represent the two symmetrical attractors while the red square is the saddle point.} 
	\label{fig: InstAndRel}%
\end{figure}

Referring to the same system defined by Eq. \eqref{eq:example_SDE}, Figure \ref{fig: bound} shows how the bounds on the transition rate in Eq. \eqref{eq BOUNDS} behave as the intensity of the shear changes through the quantity $\gamma$. A dedicated numerical method has been used to derive the three set of points that appear in Fig. \ref{fig: bound}. 
First, the Minimum Action Method (MAM) \cite{zakine2023minimum} has been implemented in Mathematica with the goal of deriving instantons - which is the main difficulty - for the chosen set of shear intensities $\gamma$. In principle, obtaining the instanton from equations Eqs. \eqref{eq HamiltonianSystem} and \eqref{eq HamiltonianSystem2} is a boundary value problem that requires to solve Eq. \eqref{eq:instanton_dyn} with, e.g., a shooting method. However, the MAM algorithm transforms the boundary value problem into an optimization problem. This is done by treating the physical time of the dynamics as an additional space coordinate, thus introducing an extra variable which plays the role of an artificial, or algorithmic  time $\tau$. Within this framework, the method is an implementation of a gradient descend on the action $\mathcal{A}$, defined in Eq. \eqref{eq pathProb}, in the space of all possible trajectories allowed by the boundary conditions:
\begin{align}
\partial_\tau \hat{\bm{x}}(t,\tau)= - \frac{\delta \mathcal{A}}{\delta \hat{\bm{x}}}, \quad  \hat{\bm{x}}(0,\tau)= \bm{x}^*, \; \hat{\bm{x}}(t_f,\tau)=\bm{x}_\nu.
\end{align}

\begin{figure}[t]
    \centering
    \includegraphics[scale=0.35]{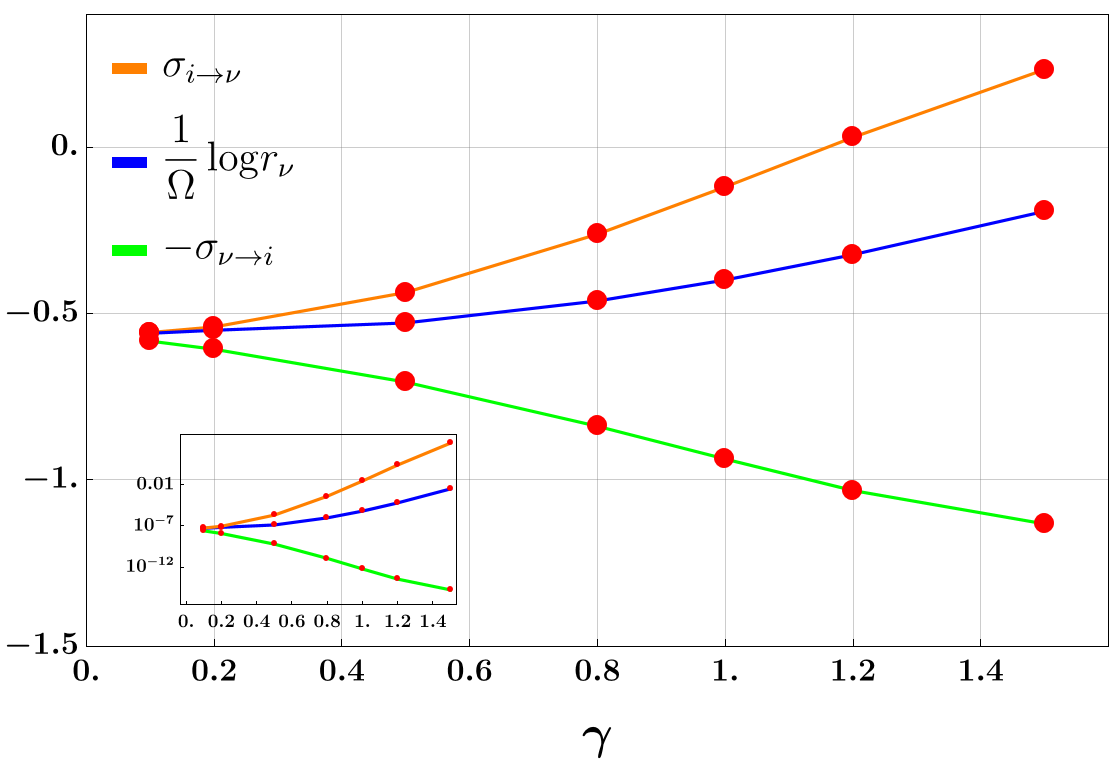}
    \caption{Bounds for the nonequilibrium transition rate in Eq. \eqref{eq BOUNDS}, at different shear intensity. Inset: bounds reported in log-scale. Data obtained with $\Omega = 30$.}
    \label{fig: bound}
\end{figure}

Having found the instanton as $\bm{x}(t)=\lim_{\tau \to \infty} \hat{\bm{x}}(t,\tau)$, one can numerically obtain the auxiliary momentum $\bm{\pi}(\bm{x})$ from equation \eqref{eq HamiltonianSystem}, as the velocity can be numerically estimated a posteriori. We showed that the instanton is the trajectory for which $\bm{\pi} = \nabla I^\text{ss}$, therefore the estimation of $\Delta I^\text{ss}$, that defines the nonequilibrium transition rate according to Eq. \eqref{eq NoneqRate}, is completed performing the integration $\Delta I^\text{ss}= \int d\bm{x}\cdot\bm{\pi}(\bm{x})$ between the stable point $\bm{x}_i$ and the saddle $\bm{x}_\nu$.
Along the instanton is then computed the upper bound in Eq. \eqref{eq BOUNDS}, where the entropy rate is defined as $\dot{\sigma} = \bm{u}(\bm{x})\cdot\dot{\bm{x}}$. The lower bound is computed integrating the same quantity along the relaxation trajectory, for each $\gamma$ value which is easily obtained via direct integration, being this dynamics given by a simpler initial value problem.

As $\gamma\to 0$, the bounds tend to saturate as expected. In that case we retrieve equilibrium,
where $\sigma_{i\to\nu} = - \sigma_{\nu\to i}$ under time reversal as expected form the odd nature of the dissipation.
}

\section{Conclusion}

In this paper, we have bridged Freidlin-Wentzell large deviation theory with nonequilibrium stochastic thermodynamics of isolated attractors, within the context of nondetailed-balance systems described by overdamped diffusion. We were able to connect the steady state rate function, known as quasipotential, to the energy and dissipation of the system. 
First, we have derived a series expansion around the detailed-balance dynamics (Eq. \eqref{eq:rf_pert_1}). It provides a powerful tool that can be directly employed numerically to evaluate $I^\text{ss}$ for nonequilibrium systems, given that it only makes use the relaxation dynamics. Concerning the numerical approach, this result suggests a straightforward iterative procedure that allows to obtain the quasipotential up to the $n^{th}$ order with a moderate effort, as relaxation dynamics can be easily implemented with well-established time integration algorithms, which are fast and relatively inexpensive. It remains to be explored how the method performs as the system dimension $d$ or the magnitude of the nonconservative forces grow. With the knowledge of the quasipotential, one can determine the escape rate $r$ from an isolated attractor according to Eq. \eqref{eq NoneqRate}. In the second part of the paper, we showed that quasipotential barriers are constrained by thermodynamics, see Eq. \eqref{eq BOUNDS}. Even though such expression consists of inequalities rather than strict equalities, this is a result in the same spirit of the Arrhenius law, given that the life-time of the attractors is linked to the energy spent to relax into and escape from it. This expression also has the advantage of relying on an experimentally accessible quantity. Moreover, for weak forcing such that the gap between the lower and upper bounds is not wide, Eq. \eqref{eq BOUNDS} becomes not just a physically relevant limit for the dynamics of the system but also a reliable predictive tool that allows us to derive a reasonable approximation of the escape rate.

\appendix

\section{Subleading correction to the rate function}\label{app:subleading}

We can leverage the instanton dynamics to obtain the subleading correction to the quasipotential in the stationary probability density $p$. In the case when the matrix $d$ is state-independent one can verify from Eq. \eqref{eq:prob_velocity} that $h_n=-d_{nm}\partial_m K$ and  Eq. \eqref{eq:ortho_ss} becomes
 \begin{align}\label{eq:correction}
 -\partial_n v_n = ( d_{nm} \partial_m I+ v_n)\partial_n K = d_t K(\bm{x}_\mathrm{ins}(t)),
 \end{align}
 where $\bm{x}_\mathrm{ins}(t)$ denotes the solution of the instanton dynamics in Eq. \eqref{eq:instanton_dyn}.
 The correction $K$, can then be obtained by integrating along the instanton trajectory (see Section \ref{sec:bounds}):
 \begin{align}
 K(\bm{x})-K(\bm{x}^*)=-\int_0^\infty  \partial_n v_n (\bm{x}_\mathrm{ins}(t)) dt.
 \end{align}
 This formula, first obtained in \cite{bouchet2016generalisation}, shows that a solenoidal macroscopic probability velocity entails no correction to the rate function. Note that $-\partial_n v_n$ is the leading order expression of the phase-space contraction rate as obtained by rewriting the Fokker-Planck equation in the form
 \begin{equation}
\frac{d}{dt} \ln p = -\partial_n j_n \underset{t \to \infty}{\longrightarrow} -\partial_n v_n + O(\epsilon),
 \end{equation}
where $d/dt= \partial_t + j_n \partial_n   $ is the material derivative, i.e., the total time derivative acting along trajectories with local mean velocity $j_n$. The phase space contraction rate plays a key role in the statistical mechanics of thermostated Hamiltonian systems \cite{evans1993probability,gallavotti1995dynamical}, corresponding to the thermodynamic entropy production rate under special conditions \cite{cohen1998note}.

\bibliographystyle{unsrt}
\bibliography{references.bib}

\end{document}


\author{Nahuel Freitas}
\affiliation{Complex Systems and Statistical Mechanics, Department of Physics and Materials Science,
University of Luxembourg, L-1511 Luxembourg, Luxembourg}
\author{Massimiliano Esposito}
\affiliation{Complex Systems and Statistical Mechanics, Department of Physics and Materials Science,
University of Luxembourg, L-1511 Luxembourg, Luxembourg}

\title{Supplementary material for ``A Maxwell demon that can work at macroscopic scales''}

\maketitle

\section{Gaussian approximation of the steady state distribution and computation of the average current}

The steady state rate function satisfies:
\begin{eqnarray}
 0 =\sum_\rho \omega_\rho(\bm{v})\left[1-e^{\bm{\Delta}_\rho\cdot \nabla f_\text{ss}(\bm{v})}\right].
 \label{eq:rf_dyn}
\end{eqnarray}
We consider the second order expansion $f_\text{ss}(\bm{v}) = \bm{v}^T C^{-1} \bm{v}/2 + \mathcal{O}(|\bm{v}|^3)$ around $\bm{v}=0$, the deterministic fixed point for $\alpha<1$. Then, from the previous equation one can show that the matrix $C$ must satisfy the Lyapunov equation
\begin{equation}
    0 = CA + A^TC + B,
    \label{eq:cov_lyapunov}
\end{equation}
where the matrices $A$ and $B$ are given by
$\{A\}_{ij} = \sum_\rho \partial_{v_i} \omega_\rho(0) (\bm{\Delta}_\rho)_j$ and $\{B\}_{ij} = \sum_\rho \omega_\rho(0) (\bm{\Delta}_\rho)_i (\bm{\Delta}_\rho)_j$
in terms of the transition rates. Eq. \eqref{eq:cov_lyapunov} can be solved explicitly for $n=1$, obtaining
\begin{equation}
\begin{split}
    C_{11} &= 
    \frac{1+ e^{\Delta V_D/2} + e^{\Delta V_S} - e^{(\Delta V_D+\Delta V_S)/2}}{2(e^{\Delta V_D/2}+e^{\Delta V_S/2}-e^{(\Delta V_D+ \Delta V_S)/2})}\\
    C_{22} &= 
    \frac{1+ e^{\Delta V_D} + e^{\Delta V_S/2} - e^{(\Delta V_D+\Delta V_S)/2}}{2(e^{\Delta V_D/2}+e^{\Delta V_S/2}-e^{(\Delta V_D+ \Delta V_S)/2})}\\
    C_{12} = C_{21} &= 
    \frac{e^{(\Delta V_D+ \Delta V_S)/2} -1}{2(e^{(\Delta V_D+\Delta V_S)/2}-e^{\Delta V_S/2}-e^{\Delta V_D/2})}.\\
\end{split}
\end{equation}
The large deviations principle $P_\text{ss}(\bm{v}) \asymp e^{-v_e^{-1} f_\text{ss}(\bm{v})}$ implies that for small $v_e$ typical fluctuations are Gaussian with covariance matrix $\mean{\bm{v}\bm{v}^T} = v_e C$. Then, to first non-trivial order in $v_e \to 0$, the average steady state current through the first inverter can
be computed as 
\begin{equation}
\begin{split}
    \mean{I_S} &\simeq \int\!\!\int d^2\bm{v} \: \frac{e^{-v_e^{-1} f_\text{ss}(\bm{v})}}{Z_\text{ss}} \: \underbrace{q_e \left(\lambda_+^\text{p/n}(\bm{v})-\lambda_-^\text{p/n}(\bm{v})\right)}_{I(\bm{v})}\\
    & \simeq I(0) + \Tr[\bm{H}_I(0) \meann{\bm{v}\bm{v}^T}]/2,
    \label{eq:macro_average_current_auto}
\end{split}
\end{equation}
where $Z_\text{ss} = \int dv \exp(-v_e^{-1} f_\text{ss}(\bm{v}))$, $\lambda_{\pm}^p(\bm{v})$ are the Poisson rates of the (p/n)MOS transistor, and $\bm{H}_I(\bm{v})$ is the Hessian of $I(\bm{v})$. In this way we can obtain the following expression for $\mean{I_S}$:
\begin{equation}
\mean{I_S} = \frac{q_e}{\tau_0}\left(e^{\Delta V_S/2}-1\right)+\frac{q_e}{\tau_0}\frac{v_e \left(2 e^{(\Delta V_D+\Delta V_S)/2}-e^{\Delta V_D+\Delta V_S/2}+e^{\Delta V_D/2+ \Delta V_S}-e^{\Delta V_D/2}+e^{\Delta V_D}-2 e^{\Delta V_S/2}\right)}{\left(e^{\Delta V_D/2}-1\right) \left(4 e^{\Delta V_S/2}\right)-4 e^{\Delta V_D/2}}+O\left(v_e^2\right),
\end{equation}
which to lower order in $\Delta V_S$ reduces to the one given in the main text.

\section{Macroscopic information flows}

The thermodynamics of measurement and feedback protocols in bipartite systems can be understood in terms of information flows which, as explained in \cite{horowitz2014}, enter as additional terms in the entropy balance of each subsystem. In this section we compute those information flows for our electronic Maxwell demon, exploiting the macroscopic limit. For this, in the sake of clarity, we will adopt a slightly different notation. We assume that we are dealing with a system with two degrees of freedom $x$
and $y$ ($v$ and $v_\text{in}$ in our example). The Poisson rates
$\lambda_\rho(x|y)$ associated to jumps $x \to x + \delta x_\rho$ depend on the variable $y$, and in the same way the Poisson rates
$\lambda_\rho(y|x)$ associated to jumps $y \to y + \delta y_\rho$ depend on the variable $x$. In our circuit, this corresponds to the fact that the Poisson rates associated to the transistors in the first inverter (which modify the voltage $v$) depend on the voltage $v_\text{in}$ (this is the feedback process in which the demon control the system), while the rates of the transistors in the second inverter (which modify $v_\text{in}$) depend on the voltage $v$ (this is the measurement process in which the demon acquires information about the state of the system). The jump sizes $\delta x_\rho$ and $\delta y_\rho$ are $\pm v_e$ in our case, with the sign depending on the transition. 

The mutual information between $x$ and $y$ is defined as $\mathcal{I} = \sum_{x,y} P_t(x,y) \log(P_t(x,y)/P_t(x)P_t(y))$ where $P_t(x,y)$ is the time-dependent global probability distribution, while $P_t(x)$ and $P_t(y)$ are the corresponding reduced distributions for $x$ and $y$. The time derivative of the mutual information accepts the following decomposition \cite{horowitz2014}:
\begin{equation}
d_t \mathcal{I} =  \mathcal{\dot I}^x + \mathcal{\dot I}^y  
\end{equation}
where 
\begin{equation}
    \mathcal{\dot I}^x = \sum_{x,y} \sum_{\rho>0} j_\rho(x|y) \log \left( \frac{P_t(y|x+\delta x_\rho)}{P_t(y|x)}\right)
    \qquad \text{and} \qquad
    \mathcal{\dot I}^y = \sum_{x,y} \sum_{\rho>0} j_\rho(y|x) \log \left( \frac{P_t(x|y+\delta y_\rho)}{P_t(x|y)}\right)
\end{equation}
are the contributions due to jumps $x\to x+\delta x_\rho$
and $y \to y + \delta y_\rho$, respectively. In the previous expressions, $j_\rho(x|y) \equiv \lambda_\rho(x|y) P_t(x,y) - \lambda_{-\rho}(x+\delta x_\rho|y) P_t(x+\delta x_\rho,y)$ is the net current along transition $x,y \to x +\delta x_\rho,y$, and $j_\rho(y|x)$ is defined similarly. Note that at steady state we have $d_t \mathcal{I} = 0$ and therefore $\mathcal{\dot I}^x = - \mathcal{\dot I}^y$. 

We will now evaluate the steady state information flows $\mathcal{\dot I}^{x/y}$ in the macroscopic limit. Under a Gaussian approximation like the one described in the previous section, the log ratio of conditional probabilities appearing in the equations above can be written as (for simplicity we assume $\mean{x}=\mean{y}=0$):
\begin{equation}
    \log\left(\frac{P_t(y|x')}{P_t(y|x)}\right) = \frac{\mean{xy}}{\mean{xx}\mean{yy}-\mean{xy}^2} 
    \left[y\:(x'-x) - \frac{\mean{xy}}{\mean{xx}} (x'^2-x^2)/2\right].
\end{equation}
Where the averages are computed for $P_t(x,y)$. 
It follows that the information flow $\mathcal{\dot I}^x$ can be 
rewritten as:
\begin{equation}
    \mathcal{\dot I}^x\:  = \frac{\mean{xy}}{\mean{xx}\mean{yy} - \mean{xy}^2} \left[ \mean{y \: d_t x}  - \frac{\mean{xy}}{\mean{xx}} \: d_t\meann{x^2} /2\right],
    \label{eq:inf_flow_gauss}
\end{equation}
where
\begin{equation}
    \mean{y \: d_t x} = \sum_{x,y} \sum_{\rho>0} j_\rho(x|y)
    \: y \: \delta x_\rho
\qquad
\text{and} 
\qquad
    d_t \meann{x^2} = \sum_{x,y} \sum_{\rho>0} j_\rho(x|y)
    \: (2  x \: \delta x_\rho + \delta x_\rho^2).
\end{equation}
The last term in Eq. \eqref{eq:inf_flow_gauss} vanishes at steady state. Then, at the Gaussian level the steady state information flows read:
\begin{equation}
    \mathcal{\dot I}_\text{ss}^x\:  = \frac{\mean{xy}  \mean{y \: d_t x}}{\mean{xx}\mean{yy} - \mean{xy}^2} 
    \qquad
    \text{and}
    \qquad
    \mathcal{\dot I}_\text{ss}^y\:  = \frac{\mean{xy}  \mean{x \: d_t y}}{\mean{xx}\mean{yy} - \mean{xy}^2}. 
    \label{eq:inf_flow_gauss_ss}
\end{equation}
Note that $\mathcal{\dot I}_\text{ss}^x + \mathcal{\dot I}_\text{ss}^x \propto \meann{x \:d_ty} + \mean{y \:d_tx} = d_t \meann{xy} = 0$ in steady state conditions, as discussed above.
A calculation similar to the one in the previous section shows that to first non-trivial order in the macroscopic limit:
\begin{equation}
    \meann{y \:d_t x} = \mean{xy} \sum_{\rho>0}\left[\partial_x \lambda_\rho(x|y)|_{0,0} - \partial_x \lambda_{-\rho}(x|y)|_{0,0} \right]\delta x_\rho 
    +\mean{yy}  \sum_{\rho>0} [\partial_y \lambda_\rho(x|y)|_{0,0} - \partial_y \lambda_{-\rho}(x|y)|_{0,0}]\delta x_\rho 
\end{equation}
Recalling that in the macroscopic limit we have the scalings 
$\lambda_\rho \propto \Omega$ and $\delta x_\rho, \mean{xx},\mean{yy},\mean{xy} \propto 1/\Omega$ with respect to the scale parameter $\Omega$, we see that $\mean{y \: d_tx} \propto 1/\Omega$ and therefore the information flows $\mathcal{\dot I}_\text{ss}^{x/y}$ are scale independent.

For the kind of bipartite systems we are considering, it was shown in \cite{horowitz2014} that when the information flows are considered in the entropy balance of each subsystem, then the following extended local second laws hold (we assume isothermal settings at temperature $T$):
\begin{equation}
\begin{split}
    \dot \Sigma_x^i &= d_t S_x + \dot \Sigma_x - k_b \mathcal{\dot I}_x\geq 0\\
    \dot \Sigma_y^i &= d_t S_y + \dot \Sigma_y - k_b \mathcal{\dot I}^y \geq 0,
    \label{eq:local_second_laws}
\end{split}
\end{equation}
where $\dot \Sigma_{x/y}^i$, $S_{x/y}$, $\dot \Sigma_{x/y}$ are, respectively, the total irreversible entropy production, the internal entropy, and the entropy flow into the environment of each subsystem. Since $S_{x/y}$ are the entropies associated to the reduced distributions $P_t(x)$ and $P_t(y)$, we see that by summing the two previous equations we recover the usual global second law $\dot \Sigma^i = \dot \Sigma^i_x + \dot \Sigma^i_y = d_t S + \dot \Sigma$, where $\dot \Sigma = \dot \Sigma_x + \dot \Sigma_y$ and $S = S_x + S_y - \mathcal{I}$ is the entropy of the full distribution $P_t(x,y)$.  Note that the local entropy flows $\dot \Sigma_{x/y}$ are proportional to $\Omega$ in the macroscopic limit, and therefore the information flows can be neglected, if everything else is fixed.

\section{Thermodynamic efficiencies for information creation and consumption}

We will now compute the information flows for the electronic Maxwell demon based on the previous result, taking $x=v$ (the output voltage of the first inverter) and $y=v_\text{in}$ (the output voltage of the second inverter). Thus, $x$ represents the system S and $y$ the demon D. Then, when evaluated at steady state, the inequalities in Eq. \eqref{eq:local_second_laws} reduce to 
\begin{equation}
\begin{split}
    \dot \Sigma_S^i &= \dot \Sigma_S + k_b \mathcal{\dot I}\geq 0\\
    \dot \Sigma_D^i &= \dot \Sigma_D - k_b \mathcal{\dot I} \geq 0,
\end{split}
\end{equation}
where $\mathcal{\dot I} = -\mathcal{\dot I}^S_\text{ss} = \mathcal{\dot I}^D_\text{ss}$ is the steady state information flow.
As we will see, $\mathcal{\dot I}$ is positive, which means that 
the dynamics of the demon produces correlations, that are consumed at the system side.

\bibliographystyle{unsrt}
\bibliography{references.bib}